\documentclass[letterpaper,11pt]{article}

\usepackage{color}
\usepackage{enumerate,latexsym}
\usepackage{amsmath,amssymb}
\makeatletter
\renewcommand*\env@matrix[1][*\c@MaxMatrixCols c]{%
  \hskip -\arraycolsep
  \let\@ifnextchar\new@ifnextchar
  \array{#1}}
\makeatother
\usepackage{graphicx}
\usepackage{amsxtra}
\usepackage{amsfonts}
\usepackage{enumerate}
\usepackage{wrapfig}
\usepackage{showkeys}
\usepackage{bm}
\usepackage{float}
\usepackage[normalem]{ulem}
\usepackage[margin=1in]{geometry}
\usepackage{hyperref}
\usepackage{fullpage}
\usepackage{fancyhdr}
\usepackage{fancyvrb}
\usepackage{lastpage}
\usepackage{color}
\usepackage{multirow}
\usepackage{relsize}
\usepackage{todonotes}
\usepackage[utf8x]{inputenc}
\usepackage{subfig}
\usepackage{dsfont}

\usepackage{amsmath}
\usepackage{fancyvrb}
\usepackage{lastpage}
\usepackage[superscript,biblabel]{cite}
\usepackage{amssymb}

\usepackage{rotating}
\usepackage{tikz}

\definecolor{b}{rgb}{.1,.1,.7}
\definecolor{rr}{rgb}{.8,0,.3}
\definecolor{g}{rgb}{0,.5,0}
\definecolor{pp}{rgb}{.5,0,.7}
\definecolor{r}{rgb}{.6,0,.3}
\definecolor{y}{rgb}{.9,.99,.9}

\newcommand{\bblock}{\begin{block}}
\newcommand{\eblock}{\end{block}}

\newcommand{\ben}{\begin{enumerate}}
\newcommand{\bit}{\begin{itemize}}
\newcommand{\een}{\end{enumerate}}
\newcommand{\eit}{\end{itemize}}
\newcommand{\ed}{\end{document}}

\begin{document}
\pagestyle{fancy}
\fvset{numbers=left,numbersep=3pt, stepnumber=5, fontsize=\scriptsize, frame=single}


\section{Introduction}
Statistical resampling methods have become feasible for parametric estimation, hypothesis testing, and model validation now that the computer is a ubiquitous tool for statisticians. This essay focuses on the resampling technique for parametric estimation known as the Jackknife procedure. To outline the usefulness of the method and its place in the general class of statistical resampling techniques, I will quickly delineate two similar resampling methods: the bootstrap and the permutation test.

\subsection{Other Sampling Methods: The Bootstrap}
The bootstrap is a broad class of usually non-parametric resampling methods for estimating the sampling distribution of an estimator. The method was described in 1979 by Bradley Efron, and was inspired by the previous success of the Jackknife procedure.\cite{efron1979bootstrap}  

Imagine that a sample of $n$ independent, identically distributed observations from an unknown distribution have been gathered, and a mean of the sample, $\bar{Y}$, has been calculated. To make inferences about the population mean we need to know the variability of the sample mean, which we know from basic statistical theory is V[$\bar{Y}$] = V[$Y$]/$n$. Here, since the distribution is unknown, we do not know the value of V[$Y$] = $\sigma^2$. The central limit theorem (CLT) states that the standardized sample mean converges in distribution to a standard normal Z as the sample size grows large|and we can invoke Slutsky’s theorem to demonstrate that the sample standard deviation is an adequate estimator for standard deviation $\sigma$ when the distribution is unknown. However, for other statistics of interest that do not admit the CLT, and for small sample sizes, the bootstrap is a viable alternative.

Briefly, the bootstrap method specifies that B samples be generated from the data by sampling with replacement from the original sample, with each sample set being of identical size as the original sample (here, $n$). The larger B is, the closer the set of samples will be to the ideal exact bootstrap sample, which is of the order of an $n$-dimensional simplex: $|C_n| = (2n-1)\bold{C}(n)$. The computation of this number, never mind the actual sample, is generally unfeasible for all but the smallest sample sizes (for example a sample size of 12 has about 1.3 million with-replacement subsamples). Furthermore, the bootstrap follows a multinomial distribution, and the most likely sample is in fact the original sample, hence it is almost certain that there will be random bootstrap samples that are replicates of the original sample. This means that the computation of the exact bootstrap is all but impossible in practice. However, Efron and Tibshirani have argued that in some instances, as few as 25 bootstrap samples can be large enough to form a reliable estimate.\cite{efron1994introduction}

The next step in the process is to perform the action that derived the initial statistic—here the mean: so we sum each bootstrap sample and divide the total by $n$, and use those quantities to generate an estimate of the variance of $\bar{Y}$ as follows:

$$
SE(\bar{Y})_B=\Big\{\frac{1}{B}\sum_{b=1}^{B}(\bar{Y}_b-\bar{Y})^2\Big\}^{1/2}
$$

The empirical distribution function (EDF) used to generate the bootstrap samples can be shown to be a consistent, unbiased estimator for the actual cumulative distribution function (CDF) from which the samples were drawn, F. In fact, the bootstrap performs well because it has a faster rate of convergence than the CLT: $O(1/n)$ vs. $O(1/\sqrt{n})$, as the bootstrap relies on the strong law of large numbers (SLLN), a more robust condition than the CLT.

\subsection{Other Sampling Methods: Permutation}
Permutation testing is done in many arenas, and a classical example is that of permuted $y$’s in a pair of random vectors $(\bold{X}, \bold{Y})$ to get a correlation coefficient p-value. For an observed sample $\bold{z} = \{(X_1, …, X_n), (Y_1, …, Y_n)\}$, the elements of (only) the $\bold{Y}$ vector are permuted B times. Then for permutation function $\pi(\cdot)$, we have that an individual permutation sample $\bold{z}_b$ is:
$$
\bold{z}_b=\{(X_1,…,X_n),(Y_{\pi(1)} ,…,Y_{\pi(n)} )\}
$$

The next step is to compute the number of times that the original correlation statistic is in absolute value greater than the chosen percentile threshold (say, 0.025 and 0.975 for an empirical $\alpha$ level of 0.05), divided by B. This value is the empirical p-value. If B = $n$! then the test is called $exact$; if all of the permutations are not performed, then there is an inflated Type I error rate, as we are less likely to sample those values in the tails of the null distribution, and hence we are less likely to say that there are values greater in absolute value than our original statistic. This method is entirely non-parametric, and is usually approximated by Monte Carlo methods for large sample sizes where the exact permutation generation is computationally impractical.\\
\\

\section{The Jackknife: Introduction and Basic Properties}
The Jackknife was proposed by M.H. Quenouille in 1949 and later refined and given its current name by John Tukey in 1956. Quenouille originally developed the method as a procedure for correcting bias. Later, Tukey described its use in constructing confidence limits for a large class of estimators. It is similar to the bootstrap in that it involves resampling, but instead of sampling with replacement, the method samples $without$ replacement.

Many situations arise where it is impractical or even impossible to calculate good estimators or find those estimators’ standard errors. The situation may be one where there is no theoretical basis to fall back on, or it may be that in estimating the variance of a difficult function of a statistic, say $g(\bar{X})$ for some function with no closed-form integral, making use of the usual route of estimation|the delta method theorem|is impossible. In these situations the Jackknife method can be used to derive an estimate of bias and standard error. Keith Knight has noted, in his book $Mathematical$ $Statistics$, that the Jackknife estimate of the standard error is roughly equivalent to the delta method for large samples.\cite{knightmathematical}\\ 

$\uline{Definition}$: The delete-1 $\bold{Jackknife}$ $\bold{Samples}$ are selected by taking the original data vector and deleting one observation from the set. Thus, there are $n$ unique Jackknife samples, and the $i$th Jackknife sample vector is defined as:

$$
\bold{X}_{[i]}=\{X_1,X_2,…,X_{i-1},X_{i+1},…,X_{n-1},X_n\}
$$
This procedure is generalizable to $k$ deletions, which is discussed further below.

The $i$th $\bold{Jackknife}$ $\bold{Replicate}$ is defined as the value of the estimator $s(\cdot)$ evaluated at the $i$th Jackknife sample. 

$$
\hat{\theta}_{(i)}:=s(\bold{X}_{[i]})
$$
The Jackknife Standard Error is defined
$$
SE(\hat{\theta})_{jack}=\Big\{\frac{n-1}{n}\sum_{i=1}^{n}(\hat{\theta}_{(i)}-\hat{\theta}_{(\cdot)})^2\Big\}^{1/2},
$$
where $\hat{\theta}_{(\cdot)}$ is the empirical average of the Jackknife replicates:
$$
\hat{\theta}_{(\cdot)}=\frac{1}{n}\sum_{i=1}^{n}\hat{\theta}_{(i)}
$$

The $(n - 1)/n$ factor in the formula above looks similar to the formula for the standard error of the sample mean, except that there is a quantity $(n - 1)$ included in the numerator. As motivation for this estimator, I consider the case that does not actually need any resampling methods: that of the sample mean. Here, the Jackknife estimator above is an unbiased estimator of the variance of the sample mean. 

To demonstrate this claim, I need to show that 
$$
\frac{n-1}{n}\sum_{i=1}^{n}(\hat{\theta}_{(i)}-\hat{\theta}_{(\cdot)})^2 = 
\frac{1}{n(n-1)}\sum_{i=1}^{n}(x_i-\bar{x})^2
$$
I note that here the Jackknife replicates in the inner squared term on the left simplify as follows:
$$
(\hat{\theta}_{(i)}-\hat{\theta}_{(\cdot)})=\frac{n\bar{x}-x_i}{n-1}-\frac{1}{n}\sum_{i=1}^{n}\bar{x}_{(i)}=\frac{1}{n-1}\Big(n\bar{x}-x_i-\frac{1}{n}\sum_{i=1}^{n}n\bar{x}-x_i\Big)=\frac{1}{n-1}(\bar{x}-x_i)
$$

Once the term is squared, the equation is complete, and is identically equal to the right hand term above. Thus, in the case of the sample mean, the Jackknife estimate of the standard error reduces to the regular, unbiased estimator commonly used. The standard error estimate is somewhat $ad$ $hoc$, but it is also intuitive. A more formal derivation was provided by Tukey and involves $pseudovalues$, which are discussed briefly below. It has been shown, however, that the Jackknife estimate of variance is slightly biased upward,\cite{efron1981jackknife} and does not work in all situations, for example, as an estimator of the median (see Knight, ibid.). In instances such as quantile estimation, it has been shown that the $\bold{delete-d}$ Jackknife, where $\sqrt{n} < d < (n - 1)$, is a consistent estimator.\cite{shao1989general}  The delete-d variance estimator has similar form as the delete-1 estimator, with a different normalizing constant: 
$$
SE(\hat{\theta})_{d-jack}=\Bigg\{\frac{n-d}{d\begin{pmatrix}n\\d\end{pmatrix}}\sum_{z}^{}(\hat{\theta}_{(z)}-\hat{\theta}_{(\cdot)})^2\Bigg\}^{1/2}
$$
The $\bold{Jackknife}$ $\bold{Bias}$ is defined as 

$$
\widehat{\text{bias}}_{jack}=(n-1)(\hat{\theta}_{(\cdot)}-\hat{\theta}),
$$
where $\hat{\theta}$ is the estimator taking the entire sample as argument. Jackknife Bias is just the average of the deviations of the replicates, which are sometimes called $Jackknife$ $Influence$ $Values$, multiplied by a factor $(n - 1)$. The bias of the sample mean is 0, so I cannot take the function $s(\cdot) = \bar{x}$ to get an idea of what the multiplier should be, as I did previously for the Jackknife SE. Instead, I consider as an estimator the uncorrected variance of the sample:
$$
\frac{1}{n}\sum_{i=1}^{n}(x_i-\bar{x})^2
$$

If I take as my estimator $\hat{\theta}$ the (biased) sample variance, I have that its bias, that is, E[$\hat{\theta}-\theta$], is equal to $-\sigma^2/n$. If I use the Jackknife bias as an estimate for the bias of my estimator, and I have that my estimator $\hat{\theta}$ is equal to the uncorrected sample variance, then the Jackknife bias formula reduces to $-S^2/n$, where $S^2$ is now the regular, corrected, unbiased estimator of sample variance. Thus, the bias here is constructed from a heuristic notion to emulate the bias of the uncorrected sample variance.

The above synopsis gave a rationale based on familiar sample-based estimators. Here is another justification: Assume that for any fixed $n$ the expected value of an estimator is the parameter estimand plus some bias term, call it $b_1(\theta)/n$. Then, as the average of the Jackknife replicates has $(n - 1)$ terms, the expected value of the average is 
$$
E[\hat{\theta}_{(\cdot)}]=\frac{1}{n}\sum_{i=1}^{n}E[\hat{\theta}_{(i)}]=\theta+\frac{b_1(\theta)}{n-1}
$$
From this observation it follows that the bias of the Jackknife replicates estimator is
$$
E[\hat{\theta}-\hat{\theta}_{(\cdot)}]=\theta+\frac{b_1(\theta)}{n}-\theta-\frac{b_1(\theta)}{n-1}=\frac{b_1(\theta)}{n(n-1)}
$$

Hence if we multiply this difference above by $(n - 1)$, we get an unbiased estimator of the bias of our original estimator. 

With the estimate of bias in hand, an obvious extension is to define a Jackknife estimate of the parameter of interest as 
$$
\hat{\theta}_{jack}=\hat{\theta}-\widehat{\text{bias}}_{jack}=\hat{\theta}-(n-1)(\hat{\theta}_{(\cdot)}-\hat{\theta})=n\hat{\theta}-(n-1)\hat{\theta}_{(\cdot)}
$$
The estimator is made clear if we remember than the Jackknife bias of the original estimator is $(n-1)(\hat{\theta}_{(\cdot)}-\hat{\theta})$, and hence the bias of the new estimator $\hat{\theta}_{jack}$  is 0. In practice, it is not always exactly 0, as the above treatment is really just a first-order Taylor series approximation, but the bias of biased estimators is often reduced by this method. If we imagine that the situation described in the explanation of the Jackknife bias where the bias is a linear combination of the estimand and a bias term were expanded so that the bias term is now an infinite series of terms, $b_1(\theta)/n + b_2(\theta)/n^2 + b_3(\theta)/n^3 + …$, and the expected value of the original estimator was that summation plus the estimand, then we have that 
$$
E[\hat{\theta}_{(\cdot)}]=\frac{1}{n}\sum_{i=1}^{n}E[\hat{\theta}_{(i)}]  =\theta+\frac{b_1(\theta)}{n-1} + \frac{b_2(\theta)}{(n-1)^2} + \frac{b_3(\theta)}{(n-1)^3} + …
$$
and
$$
E[\widehat{\text{bias}}_{jack}] =(n-1)E[\hat{\theta}-\hat{\theta}_{(\cdot)}]=
\frac{b_1(\theta)}{n} + \frac{(2n-1)b_2(\theta)}{n^2(n-1)} + \frac{(3n^2-3n+1)b_3(\theta)}{n^3(n-1)} + …
$$
and finally that the expected value of the Jackknife estimator is 
$$
E[\hat{\theta}_{jack}]=E[n\hat{\theta}-(n-1)\hat{\theta}_{(\cdot)}]=\theta-\frac{b_2(\theta)}{n(n-1)} - \frac{(2n-1)b_3(\theta)}{n^2(n-1)^2} +...\ \ \approx\ \  \theta-\frac{b_2(\theta)}{n^2}-\frac{2b_3(\theta)}{n^3}-...
$$

The above formulation shows that, as there is no first-order term $n$ in the denominators of the infinite sum terms (i.e. the first term in the Taylor expansion has cancelled out), the Jackknife bias is asymptotically smaller than the bias of any given biased estimator. \\
\\

Another construction sometimes used in Jackknife estimation is the ``pseudovalue," and can be seen as a bias-corrected version of the estimator. The scheme is to treat the jackknife pseudovalues as if they were independent random variables.

$\uline{Definition}$: The $i$th Pseudovalue of estimator $\varphi_n(\bold{X})$ for sample vector $\bold{X}$ is defined as
$$
ps_i=n\varphi_n(\bold{X})-(n-1)\varphi_{n-1}(\bold{X}_{[i]})
$$
The pseudovalues can also be written as 
$$
ps_i=\varphi_n(\bold{X})+(n-1)(\varphi_n(\bold{X})-\varphi_{n-1}(\bold{X}_{[i]})
$$
These constructs can be used in place of the replicate terms in the Jackknife SE to give confidence intervals from the $t$ distribution. However, the method is criticized by Efron and Tibshirani, who write ``This interval does not work very well: in particular, it is not significantly better than cruder intervals based on normal theory (ibid., p. 145)."

Nevertheless, it can easily be shown that pseudovalues can be used to construct a normal test of hypotheses. Since each pseudovalue is independent and identically distributed (iid), it follows that their average conforms to a normal distribution as the sample size grows large. The average of the pseudovalues is just $\hat{\theta}_{jack}$, and the expected value of that average, owing by construction to the unbiasedness of the estimator, is the parameter under investigation, $\theta$. Thus, we have that \

$$
\frac{\sqrt{n}\big(\frac{1}{n}\sum_{i=1}^{n}(n\varphi_n(\bold{X})-(n-1)\varphi_{n-1}(\bold{X}_{[i]}))-\theta\big)}{\hat{S}}\ \ \to\ \  N(0,1),
$$\

where $\hat{S}$ is the square root of the sum of the squared differences of the pseudovalue compared against $\hat{\theta}_{jack}$, divided by the sample size minus 1 (the unbiased estimate of variance). \\
\\

The Jackknife can be used in many situations. However, the method is inappropriate for correlated data or  time series data. The method assumes independence between the random variables (and identically distributed data points), and if that assumption is violated, the results will be of no use. Another condition of note is that the Jackknife estimate is composed of a linear function (subtraction) and hence will only work properly for linear functions of the data and/or parameters, or on functions that are smooth enough to be modeled as continuous without much of a problem. \\
\\

\section{Examples}
Imagine that we know that we are sampling from a uniform distribution on interval $[0, \theta]$, $\theta > 0$, and we are interested in estimating the upper bound. A simple, intuitive estimator is the sample maximum, but is this biased? The expectation of the maximum, $Y$ = max($\{X_n\}$), is 
$$
\int_{-\infty}^{\infty}y\ f(y)dy = \int_{0}^{\theta}yn\bigg(\frac{y}{\theta}\bigg)^ndy=\frac{n}{n+1}\theta
$$
This estimate is clearly biased. Since the maximum of a given fixed sample drawn from a continuous CDF is the same element for $(n - 1)$ out of the $n$ Jackknife samples, and is the second-largest term in the single Jackknife sample subset that excludes the largest element of the original sample, it is clear that the average of the Jackknife replicates is 
$$
\hat{\theta}_{(\cdot)}=\frac{n-1}{n}X_{(n)}+\frac{1}{n}X_{(n-1)}
$$
and so the Jackknife estimate of the maximum is 
$$
\hat{\theta}_{jack}=X_{(n)}+\frac{n-1}{n}(X_{(n)}-X_{(n-1)})
$$
By the results above, we have that the bias of this estimator will be smaller than that of the sample maximum, but if we had wanted to we could have just corrected the sample maximum by the constant $(n + 1)/n$ to yield a totally unbiased estimator. However, the Jackknife estimator is generalizable to $any$ distribution that has an upper bound that we would like to estimate, regardless of whether the random variable is distributed uniformly or not. This property is extremely useful when the distribution is unknown, or when it is unclear how to correct for the bias of the sample maximum.\\

To show empirically that the Jackknife bias is smaller than the maximum for $X \sim U(0, \theta)$, I wrote an \texttt{R} script that sampled from a uniform distribution on the interval $[0, 5]$, and recorded the average number of times that the absolute bias between the Jackknife estimate and 5 was less than the absolute bias between the sample maximum and 5. I ran 100,000 simulations for sample sizes 10, 30, and 100. The \texttt{R} code for sample size 100 and results follow:

\fvset{numbers=left,numbersep=3pt, stepnumber=5, fontsize=\small, frame=single}
\begin{Verbatim}
numvec<-rep(NA,100000)
maxbiasvec<-rep(NA,100000)
jackbiasvec<-rep(NA,100000)
for (i in 1:100000){
samp<-runif(100, min = 0, max = 5)
jack<-max(samp)+(100-1)/100*(max(samp)-max(samp[!samp==max(samp)]))
numvec[i]<-ifelse(abs(5-jack)<abs(max(samp)-5), 1,0)
maxbiasvec[i]<-abs(5-max(samp))
jackbiasvec[i]<-abs(5-jack)      }
mean(numvec)
mean(jackbiasvec)
mean(maxbiasvec)
\end{Verbatim}

\begin{table}[h]
\centering
\begin{tabular}{|l|lll|}
\hline
$\bold{100K}$ $\bold{Samples}$               & \multicolumn{1}{l|}{N=10} & \multicolumn{1}{l|}{N=30} & \multicolumn{1}{l|}{N=100} \\ \hline
\% JN bias $<$ Sample Max bias & 68.9\%                    & 67.73\%                   & 66.94\%                    \\  \cline{1-1}
Average Bias Jack: abs(JK-5)          & 0.4324                    & 0.1582                    & 0.0492                     \\ \cline{1-1}
Average Bias Max: abs(max-5)            & 0.4549                    & 0.1608                    & 0.0496                     \\ \cline{1-1} \hline
\end{tabular}
\end{table}\
\\
For each sample size N the Jackknife bias was smaller than the bias of the sample maximum about 68\% of the time. Of course, simply multiplying the maximum by a constant would be ideal in this case, but again, the method is generalizable to more complicated situations. \\
\\

\section{Applications}
One interesting application I came upon in a course on Microarray analysis is the genomic software application EASE (Expression Analysis Systematic Explorer).\cite{hosack2003identifying}  In experiments and analyses that use gene expression data, researchers often finalize their study with an annotated list of genes found to be differentially expressed between biological conditions. For example gene X may be expressed more in cancerous cells than in normal cells. Researchers often annotate a list of candidate genes one-by-one by looking up relevant information on the genes from an online database, or automating the process for a large gene set. The results of such a search are often difficult to interpret, especially for those not versed in the finer elements of biochemistry and cell biology. The gene set annotation will not inform nonspecialists as to whether the group of genes discovered in the analysis stage of the study is related to a plausible biological function category that has already been investigated previously: for example, genes known to regulate hemoglobin characteristics in a study to discover genes associated with sickle cell anemia.  

The EASE software queries publicly available databases with the option of adding user-defined ontology categories. The program initially gives a Fisher exact (hypergeometric) test for each known class. As an example, researchers discover gene set $\bold{A}$ is found to be associated with some phenotype of interest. They compare gene set $\bold{A}$ against annotation ontology set $\bold{X}$, which lists all known genes related to some biological function, say, apoptosis (programmed cell death) to see if gene set $\bold{A}$ is overrepresented in this biological theme more than would be expected by chance. If so, this knowledge can be used to confirm a hypothesis, or suggest a new avenue of research. Each gene in set $\bold{A}$ either is or is not in gene set $\bold{X}$, hence, we can give an exact p-value to the probability of observing at least as many genes in discovery set $\bold{A}$ as in the known ontology gene set $\bold{X}$.

The software also computes an ``EASE" score, which is a Jackknifed score for the exact test. The goal of this test is to construct a conservative score that is similar to the score given to most ontology sets, but which penalizes those known gene sets that have very few members. The program takes the discovery set $\bold{A}$, and in calculating the hypergeometric test for each ontology gene set it removes a single gene from the discovery set and computes Fisher p-values anew for every subset of genes in the discovery set having one gene excluded. \

As an example, if there is only a $single$ gene in some ontology category $\bold{X}$, and that gene happens to appear in discovery set $\bold{A}$ which contains 206 genes, then the p-value for that gene set $\bold{X}$ is 0.0152. However a different ontology gene set $\bold{Y}$ would be only slightly less significant if it had 787 genes, with 20 of our discovery set being members of that class. This is obviously a problem in commensurability. The authors of the paper write: 
\begin{quote} ``From the perspective of global biological themes, a theme based on the presence of a single gene is neither global nor stable and is rarely interesting. If the single [discovery] gene happens to be a false positive, then the significance of the dependent [ontology] theme is entirely false. However, the EASE score for these two situations is $p$ = 1 for category X and $p$ $<$ 0.0274 for category Y, and thus the EASE score eliminates the significance of the `unstable' category X while only slightly penalizing the significance of the more global theme Y." \end{quote}

\section{Differences Between Jackknife and other Resampling Methods}
That the Jackknife is asymptotically equivalent to the bootstrap is inventively shown by Efron and Tibshirani in their monograph (ibid.). They discuss the process of sampling from blocks of data as equivalent to defining a multinomial distribution on the sample n-tuple with equal probabilities for each sample point. In the case of $n$ = 3, the list of possible points is described as a triangle where each point can be selected from the set of all possible resamples of the three elements, and is graphically represented as points along the three edges of the triangle. The graphic below is reproduced from the Efron and Tibshirani paper presented in 1986.\cite{efron1986bootstrap} 

\begin{figure}[H]
\centering
\includegraphics[width=.8\linewidth]{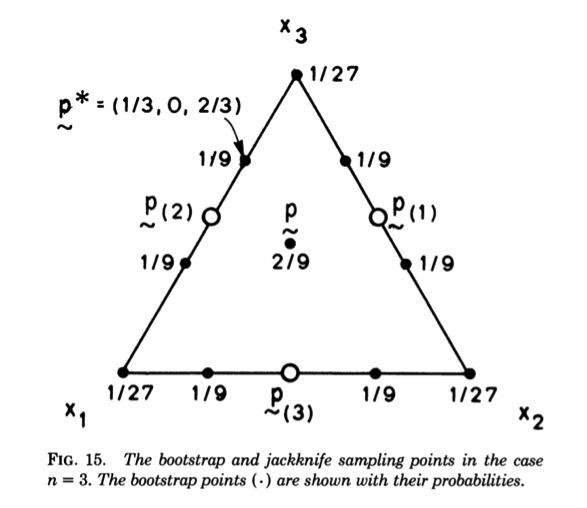}
\end{figure}

It shows the domain of the sample functional, where the ideal bootstrap sample represents the surface attained by the domain points on the simplex. Then the jackknife sample is an approximating hyperplane to the bootstrap surface. 

I now present the theorem given by Efron and Tibshirani (cf. 2, p. 287-288). 

Define $\bold{P}^0 = (1/n, …, 1/n)^T$, and $\bold{U} = (U_1, U_2, …, U_n)^T$ such that the elements of $\bold{U}$ sum to 0. Then T($\bold{P}^0$) is the original sample statistic, and T($\bold{P}_{(i)}$) is the jackknife replicate for sample point $i$: $\bold{P}_{(i)} = (1/(n-1), …, 0, … 1/(n-1))^T$.

A linear statistic T($\bold{P}^*$), where T is the functional of the vector of all probabilities such that each element is in [0, 1] and the sum of the elements is 1, has the following form:
$$
T(\bold{P}^*) = c_0 + (\bold{P}^* - \bold{P}^0)^T \bold{U}
$$
The linear statistic defines a hyperplane over simplex $S_n$. The following result states that for any statistic the jackknife estimate of the variance of T($\bold{P}^*$)  is almost the same as the bootstrap estimate for a certain linear approximation to T($\bold{P}^*$).\\

Theorem: Let $T^{LIN}$ be the unique hyperplane passing through the Jackknife points $(\bold{P}_{(i)},T(\bold{P}_{(i)}))$, $i$ = 1, 2, …, $n$. Then $var*T^{LIN} = (n - 1)/n var_{jack}$, where $var*$ is the variance under the multinomial distribution of all probability vectors. In the case $n$ = 3, it would just be the addition of the possible samples weighted by their probabilities; $var_{jack}$ is the usual formula given above, and reprinted here for clarity:
$$
\frac{n-1}{n}\sum_{i=1}^{n}(\hat{\theta}_{(i)}-\hat{\theta}_{(\cdot)})^2
$$\

Thus, the Jackknife estimate of the variance for our estimator of interest $\hat{\theta}$ is $n/(n - 1)$ times the bootstrap estimate of variance for the linear approximation to the surface described by the bootstrap simplex. \\

Proof: By solving $n$ linear equations of the form $\hat{\theta}_{(i)}= T^{LIN}(\bold{P}_{(i)})$ for $c_0$ and the components of the $\bold{U}$ vector, we find that $c_0 =\hat{\theta}_{(i)}$ and $U_i = (n - 1)(\hat{\theta}_{(\cdot)}-\hat{\theta}_{(i)} )$. Using the fact that $\bold{P}^*$ is distributed as a multiple of the multinomial distribution with mean and covariance matrix $(\bold{P}^0, [I/n^2 - \bold{P}^0 \bold{P}^0 T/n])$, and that the $U_i$ terms sum to 0 we have that $var*T^{LIN}(\bold{P}^*) = \bold{U}^T(var*\bold{P}^*)\bold{U} = 1/n^2\bold{U}^T\bold{U} = (n - 1)/n\big\{(n-1)/n\sum_{i=1}^{n}(\hat{\theta}_{(i)}-\hat{\theta}_{(\cdot)})^2\big\}$. QED.\\
\\

Thus, the accuracy of the Jackknife as a linear approximation to the bootstrap depends on how well the hyperplane $T^{LIN}$ approximates $T(\bold{P}^*)$.

\clearpage 
\bibliographystyle{ieeetr}
\bibliography{draft_bibliography.bib}

\end{document}